\documentstyle[12pt]{article}
\begin{document}
\newcommand{\nwc}{\newcommand}
\nwc{\be}{\begin{equation}}
\nwc{\ee}{\end{equation}}
\nwc{\bea}{\begin{eqnarray}}
\nwc{\eea}{\end{eqnarray}}
\nwc{\ba}{\begin{array}}
\nwc{\ea}{\end{array}}
\nwc{\rtr}{\rangle}
\nwc{\ltr}{\langle}
\nwc{\ket}[1]{|#1\rtr}
\nwc{\bra}[1]{\ltr#1|}
\nwc{\scal}[2]{\bra{#1}#2\rtr}
\nwc{\dagg}{\mbox\footnotesize{\dag}}
\nwc{\lb}{\label}
\nwc{\rf}[1]{~(\ref{#1})}
\nwc{\ci}[1]{~\cite{#1}}
\nwc{\pa}{\partial}
\nwc{\paf}[2]{\frac{\pa#1}{\pa#2}}
\nwc{\ra}{\rightarrow}
\nwc{\Tr}{\mbox{\rm Tr}}
\nwc{\de}{\delta}
\nwc{\s}{\sigma}
\nwc{\real}{\mbox{\rm Re}}
\nwc{\im}{\mbox{\rm Im}}
\nwc{\bino}[2]{\mbox{$\left(\begin{array}{c}#1\\#2\end{array}\right)$}}
\title{Ferromagnetism in the Hubbard model with an infinite-range hopping.}
\author{P. Pieri\\
{\small\it Dipartimento di Fisica, INFM and INFN, Universit\`a di Bologna,}\\
{\it\small Via Irnerio 46, I-40126, Bologna, Italy.}\\
{\small (Fax: +39 51 6305153, Tel: +39 51 6305110, E-mail: 
pieri@bologna.infn.it)}}
\date{}
\maketitle
\begin{abstract}
\noindent We prove, as recently conjectured, that the ground state of the 
Hubbard Hamiltonian with an infinite-range hopping, when the number of 
electrons $N_e=N+1$ ($N$ being the number of sites), is ferromagnetic fully 
polarized.
\end{abstract} 

\noindent Despite its long history, the problem of itinerant ferromagnetism is 
still, to many regards, an open question. The Hubbard model was 
considered from the beginning as a good model to investigate this 
intriguing problem\ci{hub}.
However, rigorous examples of ferromagnetism in the Hubbard model have 
been shown only in some rather peculiar situations: Nagaoka 
ferromagnetism for infinite Hubbard repulsion $U$ and one hole in a 
half-filled band\ci{nag} (see\ci{tas89} for a very elegant proof of 
Nagaoka's theorem), Lieb ferrimagnetism for half-filled bipartite 
lattices with sublattices containing a different number of sites
\ci{lie89} and the flat (or nearly-flat) band of Mielke\ci{mie} 
and Tasaki\ci{tas}.
Besides these rigorous results, some recent numerical and analytical 
investigations seem to indicate the existence of a fully polarized 
ferromagnetic ground state in the one dimensional Hubbard model with 
nearest and next-nearest-neighbour hopping\ci{mh,fr}. In the presence of 
so few results, it is clear that every new example of ferromagnetism in 
a Hubbard model is of great interest.\\
Very recently some conjectures about the ground state of the Hubbard 
model with an infinite-range hopping appeared in literature\ci{sal1,sal2}
(see eq.\rf{hm} below for the definition of the Hamiltonian, and 
refs.\ci{don,ver} for some previous works on this model).
Among these, one is relevant to the discussion of itinerant ferromagnetism.
According to ref.\ci{sal2}, when the filling corresponds to just one electron 
more than half-filling, the ground state of the model of Eq.\rf{hm} 
below should be a Nagaoka ferromagnetic state for every value of $U>0$. 
Now, it is easy to check that the model\rf{hm}, when $U=\infty$ and after a 
particle-hole transformation,
satisfies the conditions for the applicability of Tasaki's generalization of 
Nagaoka's theorem. 
Interestingly, the conjecture of ref.\ci{sal2} extends, for this specific 
model, the region of Nagaoka ferromagnetism  from $U=\infty$ down to 
vanishingly small $U$. In this work we shall give a rigorous proof for this 
conjecture.\\
We consider now the Hubbard Hamiltonian:
\be
H=-t \sum_{\s, 1\le i\ne j\le N} c_{i\s}^\dagger c_{j\s}
+ U \sum_{i=1}^N n_{i\downarrow} n_{i\uparrow},
\lb{hm}
\ee
The infinite-range and costant hopping in the Hamiltonian\rf{hm}, makes 
the system invariant under the action of the permutation group $S_N$ 
($N$ being the number of lattice sites). The Hamiltonian has therefore a 
block-diagonal form, with the blocks classified according to the 
irreducible representation of $S_N$. The symmetry properties under $S_N$ and 
the total spin of the ground state were studied in refs.\ci{sal1,sal2}, on 
small size systems, by exact diagonalization of the Hamiltonian\rf{hm}.
These symmetry properties and the value of the ground state's total spin were 
then supposed to hold also for systems of arbitrary size. 
In particular, while below half-filling the ground state should be a spin 
singlet, at the filling $N_e=N+1$ the ground state should be fully 
polarized: $S=(N-1)/2$ (here $S$ labels as usual the eigenvalues $S(S+1)$ 
of the total spin operator $S^2$). Finally, for filling $N_e>N+1$, the 
ground state should become completely degenerate with respect to $S$.
Hence, only when $N_e=N+1$ the ground state should be ferromagnetic (and 
fully polarized).
To find out the ground state of $H$ when $N_e=N+1$ we now need to state a 
general theorem proven by Mielke on flat-band ferromagnetism\ci{mie}.\\
Let us assume that the hopping matrix $T=(t_{ij})_{i,j=1,\ldots,N}$ of a 
Hubbard Hamiltonian $H$ has a 
lowest eigenvalue $\lambda_0$ with degeneracy $N_d$. Let 
$f^{\dagger}_{i\s}$ ($i=1,\ldots,N_d$) be the creation operator of 
one electron with spin $\s$ in one of the $N_d$ degenerate states 
with eigenvalue $\lambda_0$. Let us define:
\be
\ket{\psi}=\prod_{i=1}^{N_d}f_{i\uparrow}^{\dagger}\ket{0}
\lb{mi}
\ee
The state $\ket{\psi}$ has total spin $S=\frac{N_d}{2}$ and 
$S_z=\frac{N_d}{2}$ and is therefore fully polarized. We introduce
now the matrix $\rho$, whose elements are defined by:
\be
\rho_{jj'}=\bra{\psi}c^{\dagger}_{j\uparrow}c_{j'\uparrow}\ket{\psi}
\ee
where $j,j'=1,\ldots,N$ label the Wannier states. Let $L$ be the subset that 
contains the sites 
corresponding to the nonvanishing lines (or columns) of $\rho$.\\
{\em Theorem:} (Mielke) If and only if the matrix $\rho$ restricted to
$L$ is irreducible, then for every $U>0$, $\ket{\psi}$ is the unique ground state 
of $H$ with $N_e=N_d$ electrons, up to the trivial $S_z$ degeneracy.\\
We recall that a square matrix $A$ is called reducible if there is a 
permutation of the basis' vectors that puts it into the form:
\be
{\widetilde A}=\left(
\ba{cc}
B&0\\
C&D
\ea
\right)\ee
where $B$ and $D$ are square matrices (of course when $A$ is hermitian 
$C=0$). Otherwise $A$ is called irreducible.\\ 
Let's go back now to the Hamiltonian\rf{hm} that defines the Hubbard 
model with infinite-range hopping. We first apply a particle-hole 
transformation to $H$. The transformation, which is defined by the 
unitary operator $I=\prod_{i,\s}(c^{\dagger}_{i\s}+c_{i\s})$, transforms
some of the Fock space operators as follows:
\bea
& &Ic_{i\s}I^{\dagger}=c^{\dagger}_{i\s}\\
& &{\tilde H}\equiv IHI^{\dagger}=
t \sum_{\s, 1\le i\ne j\le N} c_{i\s}^\dagger c_{j\s}
+ U \sum_{i=1}^N n_{i\downarrow} n_{i\uparrow}+
U(N-{\hat N_e})\\
& &I{\hat N_e}I^{\dagger}=2N-{\hat N_e}\\
& &I{\vec S}I^{\dagger}=-{\vec S}
\lb{ssq}
\eea
where ${\hat N_e}$ is the particle number operator. From these relations 
it follows that if $\ket{\psi}$ is the ground state of the Hamiltonian 
${\tilde H}$ in the subspace with $N_e=N-1$ particles, then 
$I^{\dagger}\ket{\psi}$ is the ground state of $H$ in the subspace with 
$N_e=N+1$ particles. Moreover, from the relation\rf{ssq} it follows that 
if $\ket{\psi}$ is an eigenstate of $S^2$ with eigenvalue $S$, then
$I^{\dagger}\ket{\psi}$ is also an eigenstate of $S^2$ with the same 
eigenvalue. We can then look for the spin of the ground state of 
${\tilde H}$ in the subspace with $N_e=N-1$ to prove the conjecture
of ref.\ci{sal2} about the original Hamiltonian $H$.\\
The one-body part of ${\tilde H}$ is diagonalized as usual by 
introducing the Fourier-transformed operators:
$c_{k\s}=\frac{1}{\sqrt{N}}\sum_{j=1}^Ne^{ikj}c_{j\s}$ with 
$k=0,\frac{2\pi}{N},\ldots,\frac{2\pi(N-1)}{N}$. The eigenvalues of the 
hopping matrix $T$ 
are given by $E_k=t(N\de_{k,0}-1)$. The lowest eigenvalue of $T$ is 
then $-t$ and has degeneracy $N-1$. Hence, when the number of particles 
$N_e=N-1$, one of the conditions for the validity of Mielke's theorem is 
satisfied ($N_e=N_d$). The other condition to be satisfied is the 
irreducibility of the matrix $\rho$ in the subspace $L$ defined above.
In the present case the state $\ket{\psi}$ of formula\rf{mi} is given by:
\be
\ket{\psi}=\prod_{k\neq0}c_{k\uparrow}^{\dagger}\ket{0}
\ee
The matrix $\rho$ is then given by:
\bea
\rho_{jj'}&=&\bra{\psi}c^{\dagger}_{j\uparrow}c_{j'\uparrow}\ket{\psi}=
\frac{1}{N}\sum_{k,k'}\bra{\psi}
c^{\dagger}_{k\uparrow}c_{k'\uparrow}\ket{\psi}e^{i(kj-k'j')}\nonumber\\
&=&\frac{1}{N}\sum_{k,k'}\de_{k,k'}(1-\de_{k,0})e^{i(kj-k'j')}=\de_{jj'}-1/N
\eea
Since the matrix elements $\rho_{jj'}$ are all different from zero, the 
subset $L$ we defined above coincides with the whole lattice. The matrix 
$\rho$ is then irreducible since, as it follows at once from the 
definition, a necessary condition for a matrix to be reducible is to 
have at least one element equal to zero.
We conclude then that, owing to Mielke's theorem, the ground state of
${\tilde H}$ when $N_e=N-1$, is unique (up to the $S_z$ degeneracy) 
and has total spin $S=\frac{N-1}{2}$. We have therefore that, for the 
original Hamiltonian $H$, the ground state in the subspace with 
$N_e=N+1$ is unique and has total spin $S=\frac{N-1}{2}$, which is the 
conjecture we wanted to prove.\\
Wang has recently given some variational arguments in favour of this 
conjecuture\ci{wan}. His arguments are, however, not conclusive since he 
considered the stability of the 
fully polarized state with respect to a single spin-flip only.
Moreover, since it's not obvious, it should be explicitly shown that a linear 
combination of the lowest energy states of the hopping part, 
in the subspace with $S=\frac{N-3}{2}$ has always a nonzero probability to 
have two electrons on the same site, as it is argued in\ci{wan}.\\
We would like now to make some comments about Nagaoka's theorem and its 
implications for the model we have considered.  
As we said before, it's easy to check that the Hamiltonian ${\tilde H}$, 
when $U=\infty$ and $N_e=N-1$ satisfies the conditions for Tasaki's 
extension of Nagaoka's theorem. The hoppings $t_{ij}$ are $\ge 0$ as 
required by Tasaki in\ci{tas89} and the more delicate connectivity 
condition also stated by Tasaki\ci{tas89} is trivially satisfied since, in the 
present case, the hopping matrix connects directly all the lattice's 
sites. After Nagaoka's work, a central question in the theory of 
itinerant ferromagnetism has been:`` Does Nagaoka's theorem hold also 
for a finite $U$ and a small but not null hole density?"\ci{lie93}.
In this paper we have proven that for the Hamiltonian ${\tilde H}$, when
$N_e=N-1$, Nagaoka's theorem can be extended up to vanishingly small $U$.
We have no rigorous arguments to prove the conjectures of 
refs.\ci{sal1,sal2} for other filling factors, but we believe that our 
proof for the specific case $N_e=N-1$ ($N_e=N+1$ in the original 
Hamiltonian) supports also the conjectures for other fillings. In fact 
the conjectures of refs.\ci{sal1,sal2} are based on some symmetry 
arguments that seem to be independent of the filling. If we 
believe then to these conjectures, we can conclude that, when $N_e<N-1$ 
the ground state of ${\tilde H}$ is completely degenerate with respect 
to the total spin $S$. The Hamiltonian ${\tilde H}$ provides therefore a 
rather bizarre example where, with one extra hole Nagaoka's ferromagnetism 
holds up to vanishingly small $U$, while with more than one hole it disappears 
completely. Thus, this pathological example tells us, one time more, that out 
of the strict limit $U=\infty$ and one hole, every behaviour is possible and 
no definite conclusions can be drawn from Nagaoka's theorem.\\
I wish to thank D. Baeriswyl for having introduced me to the fascinating 
subject of itinerant ferromagnetism. I am also grateful to G. Morandi, E. 
Ercolessi and M. Roncaglia for helpful and stimulating discussions.

\end{document}